\documentclass[conference]{IEEEtran}
\IEEEoverridecommandlockouts
\usepackage{amsmath,amsfonts}
\usepackage{algorithmic}
\usepackage{array}
\usepackage[caption=false,font=normalsize,labelfont=sf,textfont=sf]{subfig}
\usepackage{textcomp}
\usepackage{multirow}
\usepackage{stfloats}
\usepackage{url}
\usepackage{verbatim}
\usepackage{graphicx}
\usepackage{amssymb}
\usepackage{amsthm}
\usepackage{hyperref}
\usepackage{tikz}
\usepackage{float}
\usepackage{subfig}
\usepackage{orcidlink}
\usepackage{cite}
\usepackage{caption}
\usepackage{mathrsfs}
\usepackage{cuted}
\usepackage{booktabs}
\usepackage{tabularx}   
\usepackage{booktabs}   

\usetikzlibrary{arrows.meta, positioning}
\def\BibTeX{{\rm B\kern-.05em{\sc i\kern-.025em b}\kern-.08em
		T\kern-.1667em\lower.7ex\hbox{E}\kern-.125emX}}
\usepackage{balance}
\hypersetup{
	colorlinks=true,
	linkcolor=blue,
	citecolor=blue
}
\newtheoremstyle{mystyle}
{3pt}
{3pt}
{}
{1em}
{\bfseries}
{}
{.5em}
{}

\theoremstyle{mystyle}
\newtheorem{assumptionx}{Assumption}
\newtheorem{lemmax}{Lemma}
\newtheorem{definitionx}{Definition}
\newtheorem{theoremx}{Theorem}
\newtheorem{remarkx}{Remark}

\newenvironment{definition}
{\begin{definitionx}\hspace{-0.5em}\textnormal{:}}
	{\end{definitionx}}
\newenvironment{theorem}
{\begin{theoremx}\hspace{-0.5em}\textnormal{:}}
	{\end{theoremx}}
\newenvironment{remark}
{\begin{remarkx}\hspace{-0.5em}\textnormal{:}}
	{\end{remarkx}}
\begin{document}
\title{Fixed-Time Voltage Regulation for Boost Converters\\ via Unit-Safe Saturating Functions
{
}}

\author{
    \IEEEauthorblockN{1\textsuperscript{st} Yiwei Liu}
    \IEEEauthorblockA{
        \textit{College of Westa} \\
        \textit{Southwest University} \\
        Chongqing 400715, China \\
        ORCID: 0009-0008-1283-8245
    }
    \and
    \IEEEauthorblockN{2\textsuperscript{nd} Ziming Wang}
    \IEEEauthorblockA{
        \textit{Systems Hub, Robotics and Autonomous Systems Thrust} \\
        \textit{The Hong Kong University of Science and Technology (Guangzhou)} \\
        Guangzhou 511458, China \\
        ORCID: 0000-0001-7000-9578
    }
    \and
    \IEEEauthorblockN{3\textsuperscript{rd} Xin Wang\textsuperscript{*}}
    \IEEEauthorblockA{
        \textit{College of Electronic and Information Engineering} \\
        \textit{Southwest University} \\
        Chongqing 400715, China \\
        ORCID: 0000-0003-2070-960X
    }
    \and
    \IEEEauthorblockN{4\textsuperscript{th} Yiding Ji}
\IEEEauthorblockA{
        \textit{Systems Hub, Robotics and Autonomous Systems Thrust} \\
        \textit{The Hong Kong University of Science and Technology (Guangzhou)} \\
        Guangzhou 511458, China \\
        ORCID: 0000-0003-2678-7051
    }
    \thanks{*Corresponding author: Xin Wang.}
    \thanks{This work was supported in part by the Open Foundation of the Key Laboratory of System Control and Information Processing, Ministry of Education, Shanghai, under Grant Scip20240117, and in part by the National Natural Science Foundation of China under Grant 62276214, 62303389, 62373289. Guangdong Basic and Applied Basic Research Funding grants 2022A151511076, 2024A1515012586; Guangdong Scientific Research Platform and Project Scheme grant 2024KTSCX039; Guangzhou-HKUST(GZ) Joint Funding Program grants 2024A03J0618, 2024A03J0680.}
}

	\maketitle
	
	\begin{abstract}
        This paper explores the voltage regulation challenges in boost converter systems, which are critical components in power electronics due to their ability to step up voltage levels efficiently. The proposed control algorithm ensures fixed-time stability, a desirable property that guarantees system stability within a fixed time frame regardless of initial conditions. To tackle the common chattering issues in conventional fixed-time control methods, a novel class of function families is introduced. State observers and adaptive parameters are utilized to manage the uncertainties associated with unknown load resistance. Furthermore, a new disturbance observer is developed using the proposed function family, and its advantages and limitations are illustrated through comparison with existing designs. Finally, both non-real-time and real-time simulations are conducted to validate the effectiveness and deployability of the proposed control algorithm.
	\end{abstract}

	\section{Introduction}
DC-DC boost converters play vital roles in modern power systems including electric vehicles, renewable energy, and smart grids \cite{10590071,10577254,intro000}. Their key control challenge lies in maintaining stable output voltage despite input and load variations. Effective control requires robust voltage regulation under uncertainties, fast dynamic response, and practical implementability. Advanced control strategies are therefore crucial to achieve asymptotic regulation across diverse operating conditions \cite{10438874,10268638}. In boost converter systems, the unknown load is the main source of uncertainty. To address this, disturbance observers~\cite{8036261,8584106} and adaptive parameter estimation~\cite{jia6,6213544,jia2} are commonly used. Recent advances include neural networks (NN)~\cite{10737695,jia3,jia4} and fuzzy logic systems~\cite{7021887,9951403}, which leverage universal approximation capabilities. The choice of method depends on the specific system dynamics and operating conditions.

Based on the aforementioned observation techniques, controllers can be designed for voltage regulation. PID control~\cite{bacha2014power,4801739} remains popular due to its simplicity and model-free nature. However, the boost converter's nonlinearity and non-minimum phase property (from right-half-plane zeros in linearized models) limit PID effectiveness, demanding more advanced approaches. To address boost converter control challenges, nonlinear strategies are essential. While backstepping is widely used \cite{jia7,10274473,jia1,jia5}, it requires system transformation to meet strict-feedback form requirements. Among its variants, fixed-time control \cite{9841443,10233088} has gained attention for providing convergence independent of initial conditions, unlike finite-time methods \cite{10164741,10734366} whose performance varies with operating points. Although predefined-time control \cite{11048942} enables exact convergence-time specification, its time-varying gains increase system complexity, making fixed-time control a balanced compromise.

However, fixed-time approaches face two key limitations: (1) chattering from discontinuous sign functions in nonlinear gains, and (2) singularity issues from terms like $e^{a-1}$ in virtual control laws. While smooth alternatives (e.g., hyperbolic tangent \cite{10638813}) offer partial solutions, they lack universal applicability. This highlights the need for more generalized function families to enhance fixed-time control's practicality in boost converter applications.

The main contributions are summarized as follows: (1) The most significant contribution of this work is the introduction of a novel class of function families that can be employed in controller design to completely eliminate both singularity and chattering issues. It is shown that the commonly used hyperbolic tangent function is merely a special case of the proposed function class. Furthermore, a general scaling theorem is established based on this function family, which provides a rigorous foundation for the control design process. (2) Leveraging the proposed function family, a fixed-time control scheme is developed to ensure that, under arbitrarily bounded disturbances, the voltage tracking error converges to a neighborhood of the reference voltage within a fixed time. In addition, a new class of fixed-time observers is designed using the same function family, thereby further enhancing the control performance of the overall system.

The structure of this paper is organized as follows. Section~\ref{Sec3} presents the boost converter modeling, control objectives formulation, and introduces the novel function families that form the basis of our proposed control scheme. Section~\ref{Sec4} validates the method through comparative simulations and real-time implementation results, demonstrating both superiority and practical deployability. Finally, Section~\ref{Sec5} concludes the paper and discusses future research directions.

	\begin{figure}[!t]
		\centering
		\includegraphics[width=1\linewidth]{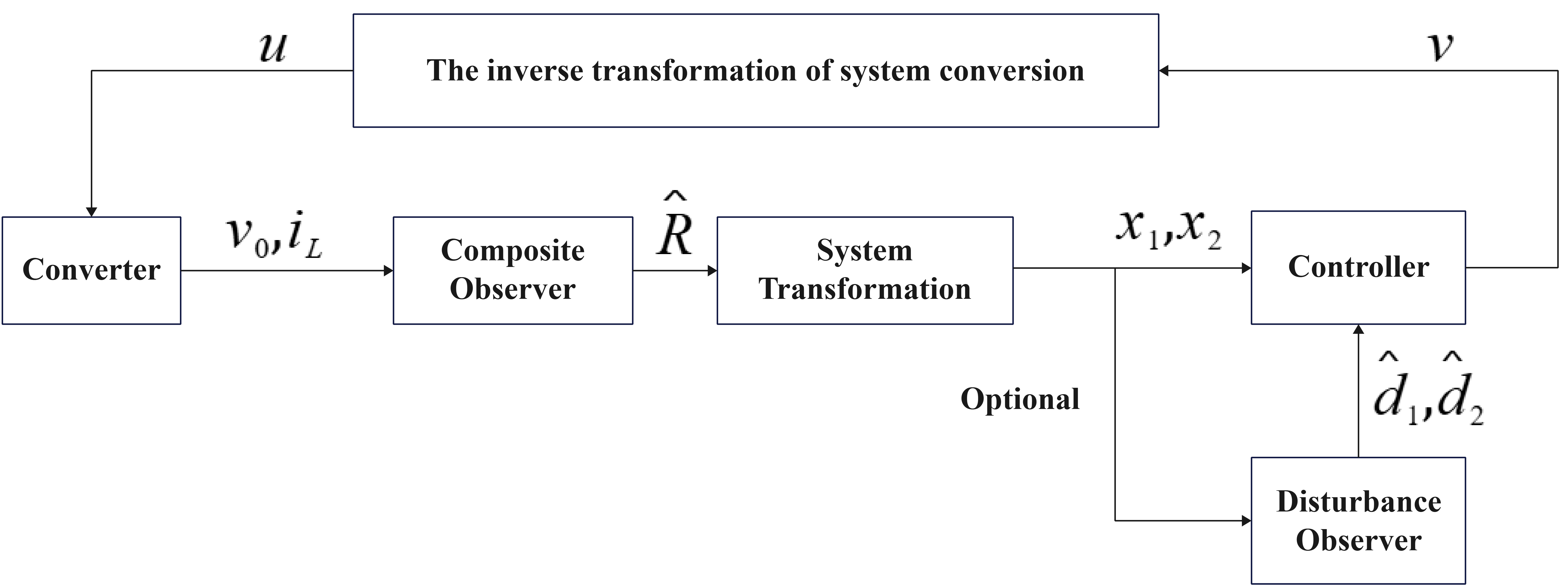}
		\caption{The proposed control framework.}
		\label{Fig.-1}
	\end{figure}
    
	\section{Main Results}\label{Sec3}
	\subsection{Unit-Safe Saturating Function}
	First, we begin by defining a special class of functions.
	\begin{definition}
		A map \(f:\mathbb{R}\to(-1,1)\) is called a \emph{Unit-Safe Saturating Function} (USSF) if and only if it satisfies the following properties:
		\begin{enumerate}
			\item \textbf{Smoothness:} \(f\in C^{\infty}(\mathbb{R})\).
			\item \textbf{Oddness:} \(f(-x)=-f(x)\) for every \(x\in\mathbb{R}\).
			\item \textbf{Strict Monotonicity:} \(f'(x)>0\) for every \(x\in\mathbb{R}\).
			\item \textbf{Saturation:} \(\displaystyle\lim_{x\to\pm\infty}f(x)=\pm1\).
			\item \textbf{Slope-limited safety:} \(\displaystyle x^{2}f'(x)\le \varepsilon \quad \forall\,x\in\mathbb{R}\), where $\varepsilon$ is a positive constant.
		\end{enumerate}
        
		The set of all such functions is denoted by \(\mathscr{F}_{\mathrm{USSF}}\).
	\end{definition}
	
	\begin{theorem}\label{thm:ussf_bound}
		Let \(f\in\mathscr{F}_{\mathrm{USSF}}\) be an arbitrary USSF and define
		$
		g(x)=|x|-x\,f(x),
		M=\sup_{x\in\mathbb{R}}g(x).
		$
		Then
		$
		M\leq\varepsilon.
		$
	\end{theorem}
	\begin{IEEEproof}
		Observe first that, owing to the oddness of \(f\), the function \(g\) defined above is necessarily even. Consequently, the analysis can be confined without loss of generality to the domain \(x \geq 0\), wherein the expression for \(g(x)\) simplifies to
		\begin{equation}
			g(x) = x(1 - f(x)).
		\end{equation}
        
		Now, invoking the asymptotic property \(\lim_{t \to \infty} f(t) = 1\), and applying the Fundamental Theorem of Calculus, one can equivalently express the quantity \(1 - f(x)\) in integral form as $1 - f(x) = \int_x^\infty f'(t)\,dt$. On the other hand, by virtue of the slope-limited safety condition, which imposes the pointwise bound \(t^2 f'(t) \leq \varepsilon\) for all \(t > 0\), it has $f'(t) \leq \frac{\varepsilon}{t^2}$. Then, substituting this upper bound into the integral representation yields $1 - f(x) \leq  \frac{\varepsilon}{x}$. Consequently, the function $g(x)$ admits the upper bound
		\begin{equation}
			g(x) = x(1 - f(x)) \leq x \cdot \frac{\varepsilon}{x} = \varepsilon, \quad \forall\, x > 0.
		\end{equation}
        
		In particular, evaluating $g(x)$ at the origin satisfies $g(0) = 0 \leq \varepsilon$. While for negative arguments, the evenness of \(g\) guarantees that $g(x) = g(-x) \leq \varepsilon$ with $\forall x<0$.
		Combining these observations, we deduce that \(g(x) \leq \varepsilon\) for all \(x \in \mathbb{R}\). Then, taking the supremum over \(\mathbb{R}\), we get
		$M = \sup_{x \in \mathbb{R}} g(x) \leq \varepsilon$.
	\end{IEEEproof}

\begin{figure}[!t]
		\centering
		\includegraphics[width=1\linewidth]{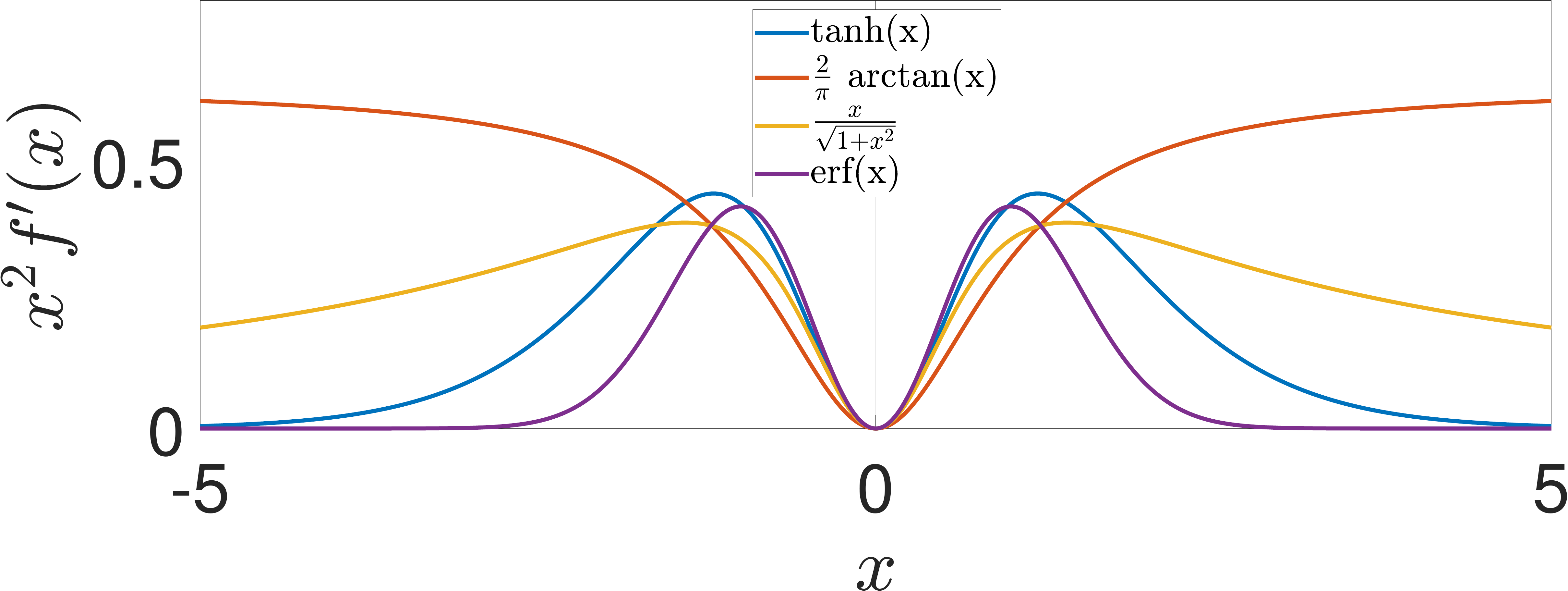}
		\caption{The plots of $x^{2}f'(x)$.}
		\label{Fig.0}
	\end{figure}

    \begin{remark}
        In fact, it can be observed that the commonly used functions $\tanh(x)$, $\frac{2}{\pi}\arctan(x)$, and $\frac{x}{\sqrt{1+x^{2}}}$ all satisfy the conditions of a $\mathrm{USSF}$. Moreover, according to the formal definition of $\mathrm{USSF}$, functions such as $\mathrm{erf}(x)$ also fall into this category. Therefore, the class of $\mathrm{USSF}$ functions encompasses a broader family of smooth saturating nonlinearities, which can be extensively utilized in the design of control laws. To provide a more intuitive illustration, Fig.~\ref{Fig.0} presents the plots of $x^{2}f'(x)$ for $f(x) = \tanh(x)$, $\frac{2}{\pi}\arctan(x)$, $\frac{x}{\sqrt{1+x^{2}}}$, and $\mathrm{erf}(x)$, respectively. In addition, Table.~\ref{Table1} reports the corresponding precise values of $\varepsilon$ for each case.
    \end{remark}
	 
	\begin{table}[!ht]
		\centering
		\begin{tabular}{c c}
			\toprule
			Function $f(x)$ & $\varepsilon$ \\ \midrule
			$\tanh x$ & 0.4392288 \\
			$\displaystyle \dfrac{2}{\pi}\arctan x$ & 0.6366198 \\
			$\displaystyle \dfrac{x}{\sqrt{1+x^{2}}}$ & 0.3849002 \\
			$\operatorname{erf}x$ & 0.4151075 \\ \bottomrule
		\end{tabular}
		\caption{Slope–limit constants $\varepsilon=\max_{x\in\mathbb R}x^{2}f'(x)$ for four representative USSFs.}
		\label{Table1}
	\end{table}

    \subsection{Problem Formulation}
	The dynamic equations of a boost converter is designed as
	\begin{equation}
		\begin{aligned}
			\frac{dv_0}{dt} &= \frac{1 - u}{C} i_L - \frac{v_0}{RC}, \\
			\frac{di_L}{dt} &= -\frac{1 - u}{L} v_0 + \frac{V_i}{L}, \label{eq1}
		\end{aligned}
	\end{equation}
	where $v_0$ denotes the capacitor voltage (i.e., the output voltage), $i_L$ represents the inductor current, and $V_i$ is the input voltage. The parameter $R$ stands for the load resistance, $C$ is the nominal capacitance, $L$ is the nominal inductance, and $u \in [0,1]$ denotes the duty ratio input. In this context, $R$ is assumed to be unknown. The reference trajectory of the output voltage is denoted by $v_r$. Accordingly, the control objective is to design the duty ratio input $u$ such that the output voltage $v_0$ effectively tracks the reference voltage $v_r$.

	\subsection{Control Law Design}
	Since the dynamic equations of the boost converter are not in standard form, which is unfavorable for backstepping design, by employing the diffeomorphic transformation proposed in \cite{8036261,11048942}, the system can be reformulated as
\begin{equation}
	\begin{aligned}
		\dot{x}_1 &= x_2 + d_1, \\
		\dot{x}_2 &= \nu + d_2, \label{eq4}
	\end{aligned}
\end{equation}
where the transformed state variables and associated terms are defined as follows:
\begin{align*}
	x_1 &= \frac{1}{2}\left(Cv_0^2 + Li_L^2\right), \\
	x_2 &= V_i i_L - \frac{v_0^2}{\hat{R}}, \\
	\nu &= \frac{V_s^2}{L} + \frac{2v_0^2}{\hat{R}^2 C} - \left(\frac{V_s v_0}{L} + \frac{2i_L v_0}{\hat{R} C}\right)(1 - u), \\
	d_1 &= \frac{v_0^2}{\hat{R}} - \frac{v_0^2}{R}, \\
	d_2 &= \frac{2}{\hat{R}C} \left( \frac{v_0^2}{R} - \frac{v_0^2}{\hat{R}} \right).
\end{align*}

Here, $\hat{R}$ denotes the estimated value of $R$, which will be defined in subsequent sections.
	The corresponding reference trajectory is given by
	\begin{equation}
		x_r = \frac{L}{2V_i^2} \left( \frac{v_r^2}{R} \right)^2 + \frac{1}{2} C v_r^2.
	\end{equation}
    
	Thus, the control objective is to design the input $\nu$ such that $x_1$ asymptotically converges to $x_r$. First, since $R$ is unknown, the reference trajectory $x_r$ cannot be directly utilized. Therefore, an observer must be designed to estimate $R$. According to the method proposed in \cite{6213544}, we design the following observer:
	
	\begin{equation}
		\begin{aligned}
			\dot{\hat{i}}_L &= -(1 - u)\frac{\hat{v}_0}{L} + \frac{V_s}{L} + K_1(i_L - \hat{i}_L), \\ 
			\dot{\hat{v}}_0 &= (1 - u)\frac{\hat{i}_L}{C} - \frac{1}{\hat{R}C}v_0 + K_2(v_0 - \hat{v}_0), \label{observer}
		\end{aligned}
	\end{equation}
	where $\hat{i}_L$ is the estimated inductor current, $\hat{v}_0$ is the estimated output voltage, $\hat{R}$ denotes the estimated load resistance, and $K_1$ and $K_2$ are observer gains. For notational convenience, define the conductance as \( G = \frac{1}{R} \). The estimated conductance \( \hat{G} \) is then updated according to the following adaptive law:
\begin{equation}
\dot{\hat{G}} = -\kappa v_0 \tilde{v}_0,
\end{equation}
	where $\tilde{v}_0 = v_0 - \hat{v}_0$ denotes the output voltage estimation error, and $\kappa$ is a user-defined positive constant. According to the results established in \cite{6213544}, it can be shown that $\hat{v}_0$, $\hat{i}_L$, and $\hat{G}$ asymptotically converge to $v_0$, $i_L$, and $G$, respectively. Consequently, the disturbance terms $d_1$ and $d_2$ also asymptotically converge to zero. Accordingly, we define a new implementable reference trajectory $\hat{x}_r$ as
	\begin{equation}
		\hat{x}_r = \frac{L}{2V_i^2} \left( \frac{v_r^2}{\hat{R}} \right)^2 + \frac{1}{2} C v_r^2.
	\end{equation}
	
	We define the tracking errors as follows:
	\begin{equation}
		\begin{aligned}
			e_1 &= x_1 - \hat{x}_r, \\
			e_2 &= x_2 - \alpha,
		\end{aligned}\label{control law}
	\end{equation}
	where $\alpha$ denotes a virtual control input. According to the design principles of the backstepping method, the control strategy is formulated as
    \begin{equation}
		\begin{aligned}
			\alpha &= -k_1 f_1 - k_2 e_1^{\iota - 1} f_2 - k_3 e_1+ \dot{\hat{x}}_r, \\ 
			\nu &= -k_4 g_1 - k_5 e_2^{\iota - 1} g_2 - k_6 e_1+ \dot{\alpha},
		\end{aligned}
	\end{equation}
	where $k_1$, $k_2$, $k_3$, $k_4$, $k_5$, and $k_6$ are positive control gains to be selected by the user, and $\iota > 2$ is also a user-defined design parameter. The functions $f_1 = f(e_1)$, $f_2 = f(e_1^\iota)$, $g_1 = g(e_2)$, and $g_2 = g(e_2^\iota)$ are constructed using $\mathrm{USSF}$ functions. The forms of $f(\cdot)$ and $g(\cdot)$ may be chosen identically or independently, depending on specific implementation requirements. The analytical expression for $\dot{\alpha}$ is given by
\begin{equation}
\begin{aligned}
\dot{\alpha} ={}& (x_2 + d_2 - \dot{\hat{x}}_r) \Big( 
    -k_1 \frac{\partial f_1}{\partial e_1} 
    - k_2 (\iota - 1) e_1^{\iota - 2} f_2 \\
&\quad - k_2 \iota e_1^{2\iota - 2} \frac{\partial f_2}{\partial (e_1^\iota)} 
    - k_3 \Big) + \ddot{\hat{x}}_r.
\end{aligned}
\end{equation}

It is evident that the presence of the unknown term $d_2$, the exact value of $\dot{\alpha}$ cannot be computed analytically.Fortunately, there already exist a wide range of accurate numerical methods to approximate $\dot{\alpha}$, such as dynamic surface control (DSC) \cite{880994} and fixed-time differentiators \cite{9395197}. Alternatively, if numerical differentiation is not adopted, one may substitute $\dot{\alpha}$ in the control input $\nu$ with its estimate $\dot{\hat{\alpha}}$, given by:
\begin{equation}
\begin{aligned}
\dot{\hat{\alpha}} ={}& (x_2 - \dot{\hat{x}}_r) \Big( 
    -k_1 \frac{\partial f_1}{\partial e_1} 
    - k_2 (\iota - 1) e_1^{\iota - 2} f_2 \\
&\quad - k_2 \iota e_1^{2\iota - 2} \frac{\partial f_2}{\partial (e_1^\iota)} 
    - k_3 \Big) + \ddot{\hat{x}}_r.
\end{aligned}
\end{equation}

Since $d_2$ asymptotically converges to zero, it follows that $\dot{\hat{\alpha}}$ also asymptotically converges to $\dot{\alpha}$.

	\begin{theorem}
		For the control scheme given in \eqref{control law} and the observer defined in \eqref{observer}, it can be proven that the closed-loop system exhibits semi-global fixed-time stability (SPFTS), i.e., the output voltage $v_0$ converges to a neighborhood of the reference voltage $v_r$ within a fixed time.
	\end{theorem}
	\begin{IEEEproof}
Consider the following Lyapunov function:
$
	V = \frac{1}{2}e_1^2 + \frac{1}{2}e_2^2.
$
Under the condition $k_3, k_6 > \tfrac{1}{2}$, application of \eqref{control law} with Theorem~\ref{thm:ussf_bound} yields:
\begin{equation}
\dot{V} \leq -\kappa_1 V^{\frac{1}{2}} - \kappa_2 V^{\frac{\iota}{2}} + C
\end{equation}
where $\kappa_1 = 2\min(k_1,k_4)$, $\kappa_2 = 2\min(k_2,k_5)$, and $C = \tfrac{1}{2}(\bar{d}_1^2 + \bar{d}_2^2) + (k_1+k_2+k_4+k_5)\varepsilon$. According to \cite[Lemma 2]{10233088}, the system satisfies the conditions for SPFTS.
	\end{IEEEproof}
    \begin{remark}
    It follows that for all disturbances with known upper bounds, the proposed control scheme guarantees that the tracking errors converge to a compact set within a fixed time, where the size of the compact set is determined by the magnitude of the disturbance bounds. In particular, the terms $d_1$ and $d_2$ arise solely due to the uncertainty in the exact value of the load resistance $R$. In other words, if the exact value of $R$ were known, we would have $d_1 = d_2 = 0$. Moreover, as previously analyzed, $d_1$ and $d_2$ asymptotically converge to zero. Therefore, the compact set to which the voltage tracking error converges continues to shrink over time, eventually reducing to the minimal invariant set when $d_1 = d_2 = 0$.
    \end{remark}

\subsection{Disturbance Observer Design}

Since the transformed system essentially exhibits a strict-feedback structure, it is possible to design observers to estimate $d_1$ and $d_2$, thereby enabling further enhancement of the overall control performance. The disturbance observer is constructed as follows:
\begin{equation}
\begin{aligned}
\dot{\hat{x}}_1 &= x_2 + \hat{d}_1, \\
\hat{d}_1 &= -\kappa_1 r_1 - \kappa_2 \tilde{x}_1^{\theta - 1} r_2 - \kappa_3 \tilde{x}_1, \\
\dot{\hat{x}}_2 &= \nu + \hat{d}_2, \\
\hat{d}_2 &= -\kappa_4 h_1 - \kappa_5 \tilde{x}_2^{\theta - 1} h_2 - \kappa_6 \tilde{x}_2,
\end{aligned}
\label{disturbance_observer}
\end{equation}
where $\kappa_1$ through $\kappa_6$ are user-defined positive observer gains, and $\theta > 2$ is a design parameter. The functions $r(\cdot)$ and $h(\cdot)$ belong to the class of USSF. Specifically, with the state estimation errors $\tilde{x}_1 = \hat{x}_1 - x_1$ and $\tilde{x}_2 = \hat{x}_2 - x_2$, we have
$r_1 = r(\tilde{x}_1), r_2 = r(\tilde{x}_1^\theta), h_1 = h(\tilde{x}_2), h_2 = h(\tilde{x}_2^\theta)$.

\begin{theorem}
Define the disturbance estimation errors as $\tilde{d}_1 = \hat{d}_1 - d_1$ and $\tilde{d}_2 = \hat{d}_2 - d_2$. The proposed disturbance observer in \eqref{disturbance_observer} guarantees that $\tilde{d}_1$ and $\tilde{d}_2$ converge to compact sets within a predefined fixed time.
\end{theorem}

\begin{IEEEproof}
Consider the Lyapunov function $L_1 = \frac{1}{2} \tilde{x}_1^2$. Taking its time derivative, substituting \eqref{disturbance_observer}, and applying Young's inequality, with $\bar{d}_1$ as an upper bound of the disturbance, $\varepsilon$ as the residual of the USSF approximation, we obtain:
\begin{equation}
\dot{L}_1 \leq -2\kappa_1 L_1^{\frac{1}{2}} - 2\kappa_2 L_1^{\frac{\theta}{2}} + \kappa_1 \varepsilon + \kappa_2 \varepsilon + \frac{1}{2} \bar{d}_1^2,
\end{equation}

By \cite[Lemma 2]{10233088}, this inequality implies that $\tilde{x}_1$ converges to a compact set within a predefined fixed time. Since $\tilde{d}_1$ is a smooth function of $\tilde{x}_1$, it also converges to a compact set in fixed time. A similar argument applies to $\tilde{d}_2$.
\end{IEEEproof}

If the proposed disturbance observer is employed, with the observer generating disturbance estimates $\hat{d}_1$ and $\hat{d}_2$, the control laws are modified as follows:
\begin{equation}
		\begin{aligned}
			\alpha &= -k_1 f_1 - k_2 e_1^{\iota - 1} f_2 - k_3 e_1 - \hat{d}_1 + \dot{\hat{x}}_r,\\ 
			\nu &= -k_4 g_1 - k_5 e_2^{\iota - 1} g_2 - k_6 e_1 - \hat{d}_2 + \dot{\alpha},
		\end{aligned}
	\end{equation}

\begin{remark}
While employing a disturbance observer can significantly improve the overall control performance, it also introduces additional tunable parameters and increases the complexity of the control architecture. This added complexity may result in a higher computational burden. Therefore, the decision to incorporate the observer should be based on the trade-off between control performance and implementation cost, particularly in resource-constrained embedded systems.
\end{remark}

	\section{Simulation Verification}\label{Sec4}
	\subsection{Simulation Test}
First, we conduct simulation tests to preliminarily verify the effectiveness of the proposed scheme. The parameters of the boost converter are set as follows: $L=10\,\mu\mathrm{H}$, $C=100\,\mu\mathrm{F}$, $V_i=6\,\mathrm{V}$. The load resistance is designed as
	\[
	R=\begin{cases}
		10\,\Omega, & \text{if } t \in (0, 0.2\,\text{s}) \\
		20\,\Omega, & \text{if } t \in [0.2\,\text{s}, 0.6\,\text{s}) \\
		10\,\Omega, & \text{if } t \in [0.6\,\text{s}, 1\,\text{s}].
	\end{cases}
	\]
    
	The switching frequency is $f_s=100\,\mathrm{kHz}$. The reference voltage is set to $v_r=12\,\mathrm{V}$. The control gains are chosen as $k_1=k_2=10000, k_3=1$ and $k_4=k_5=90000, k_6=1$. The exponent parameter is chosen as
	$\iota = 3.$ The observer gains are selected as
	$K_1 = K_2 = 4165, \quad \gamma = 200.$ The USSF function we chose is 
\( f(\cdot) = g(\cdot) = \frac{\cdot}{\sqrt{1 + (\cdot)^2}}\)
. To reduce computational load and simplify the parameter tuning process, the disturbance observer is not employed in the control design. 

	To further illustrate the benefits of the proposed control scheme, we conduct a comparative analysis with both a cascaded PID controller and the control strategy outlined in \cite{8036261}. The boost converter utilizes a cascaded control configuration comprising two hierarchically organized feedback loops \cite{bacha2014power}:
\begin{equation*}
	\begin{cases}
		\alpha(t) = \kappa_v^{(p)} e_1(t) + \kappa_v^{(i)} \displaystyle\int_0^t e_1(\tau) d\tau + \kappa_v^{(d)} \dfrac{de_1(t)}{dt}, \\[1.5ex]
		u(t) = \kappa_i^{(p)} e_2(t) + \kappa_i^{(i)} \displaystyle\int_0^t e_2(\tau) d\tau + \kappa_i^{(d)} \dfrac{de_2(t)}{dt},
	\end{cases}
\end{equation*}
where the gain values are given as \( \kappa_v^{(p)} = 5 \), \( \kappa_v^{(i)} = 40 \), \( \kappa_v^{(d)} = 0 \), and \( \kappa_i^{(p)} = 20 \), \( \kappa_i^{(i)} = 1 \), \( \kappa_i^{(d)} = 0 \), respectively, representing the proportional, integral, and derivative parameters of the voltage and current controllers.

	The controller proposed in \cite{8036261} is designed as follows:
	\begin{equation*}
		\alpha = -c_1 e_1  + \dot{\hat{x}}_r,
	\end{equation*}
	\begin{equation*}
		\nu = -c_2 e_2 + \ddot{\hat{x}}_r,
	\end{equation*}
where $c_1=c_2=100000$.

	To facilitate a more comprehensive performance comparison, we introduce the following error metrics: Root Mean Square Error (RMSE), Mean Square Error (MSE), and Mean Absolute Error (MAE).
	
	\begin{figure}[!t]
		\centering
		\includegraphics[width=1\linewidth]{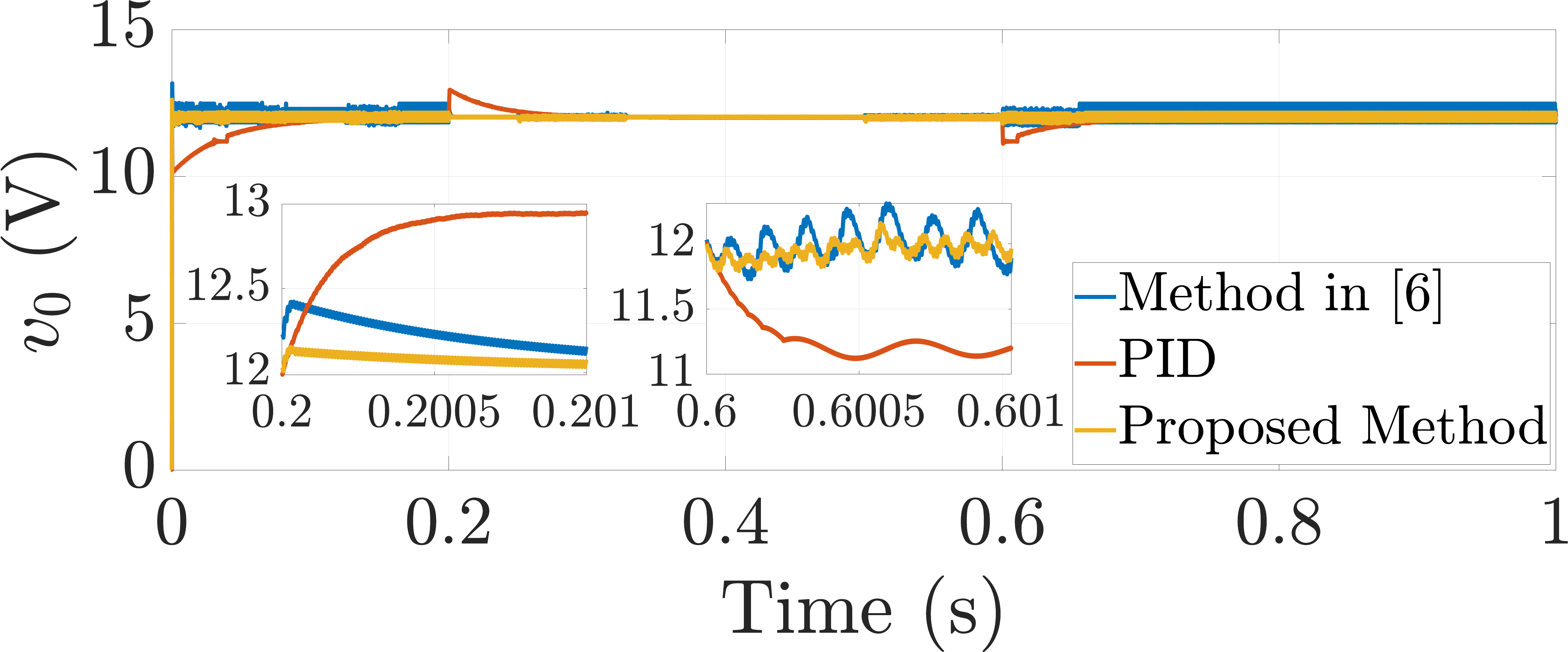}
		\caption{The output voltage response.}
		\label{fig2}
	\end{figure}
	
	\begin{figure}[!t]
		\centering
		\includegraphics[width=1\linewidth]{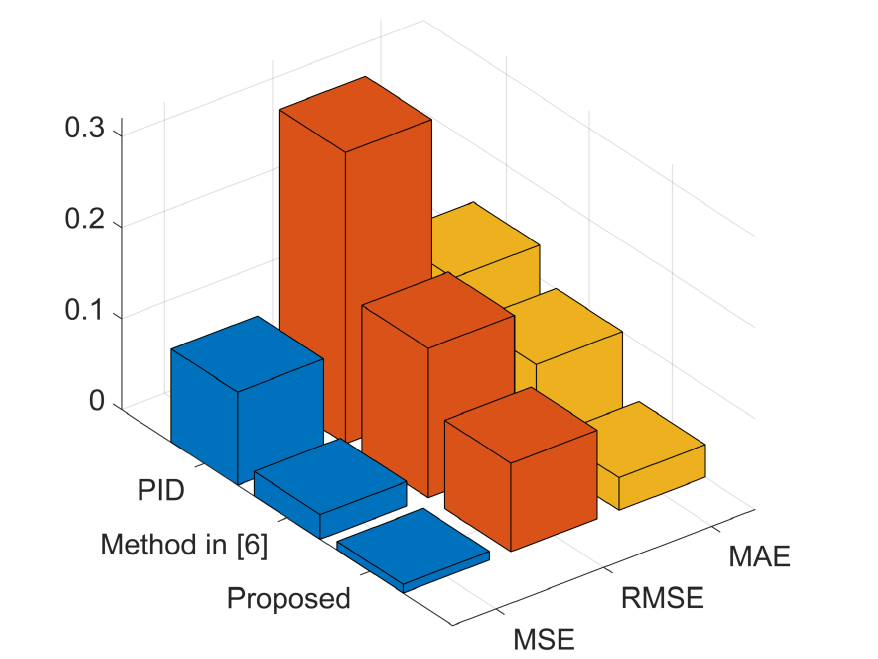}
		\caption{Error metrics.}
		\label{fig3}
	\end{figure}
	
	\begin{table}[htbp]
		\centering
		\caption{Performance comparison of different methods}
		\begin{tabular}{lccc}
			\hline
			\textbf{Method} & \textbf{MSE} & \textbf{RMSE} & \textbf{MAE} \\
			\hline
			Method in \cite{8036261}      & 0.026878  & 0.163946  & 0.100643 \\
			PID                 & 0.102183  & 0.319661  & 0.136390 \\
			Proposed Method     & 0.009420  & 0.097056  & 0.035683 \\
			\hline
		\end{tabular}
		\label{tab:error_metrics}
	\end{table}

	The simulation results are presented in Fig.~\ref{fig2} and Fig.~\ref{fig3}. As illustrated in Fig.~\ref{fig2}, all three methods are capable of stabilizing the output voltage around 12\,V. Furthermore, when a sudden change in the load resistance occurs, each method either maintains the voltage near 12\,V or drives it back to 12\,V after a transient deviation. 
	
	A closer examination of the subplots reveals that the proposed method achieves significantly faster convergence following abrupt changes in the load, leading to quicker restoration of the reference voltage compared to the other two methods. To better visualize the performance in terms of error, Fig.~\ref{fig3} presents the RMSE, MSE, and MAE metrics for all three methods, while Table.~\ref{tab:error_metrics} reports their precise numerical values. These results confirm that the proposed control strategy indeed exhibits superior voltage regulation performance.

	\subsection{Real-time Simulation Results}
\begin{figure}[!t]
		\centering
		\includegraphics[width=0.8\linewidth]{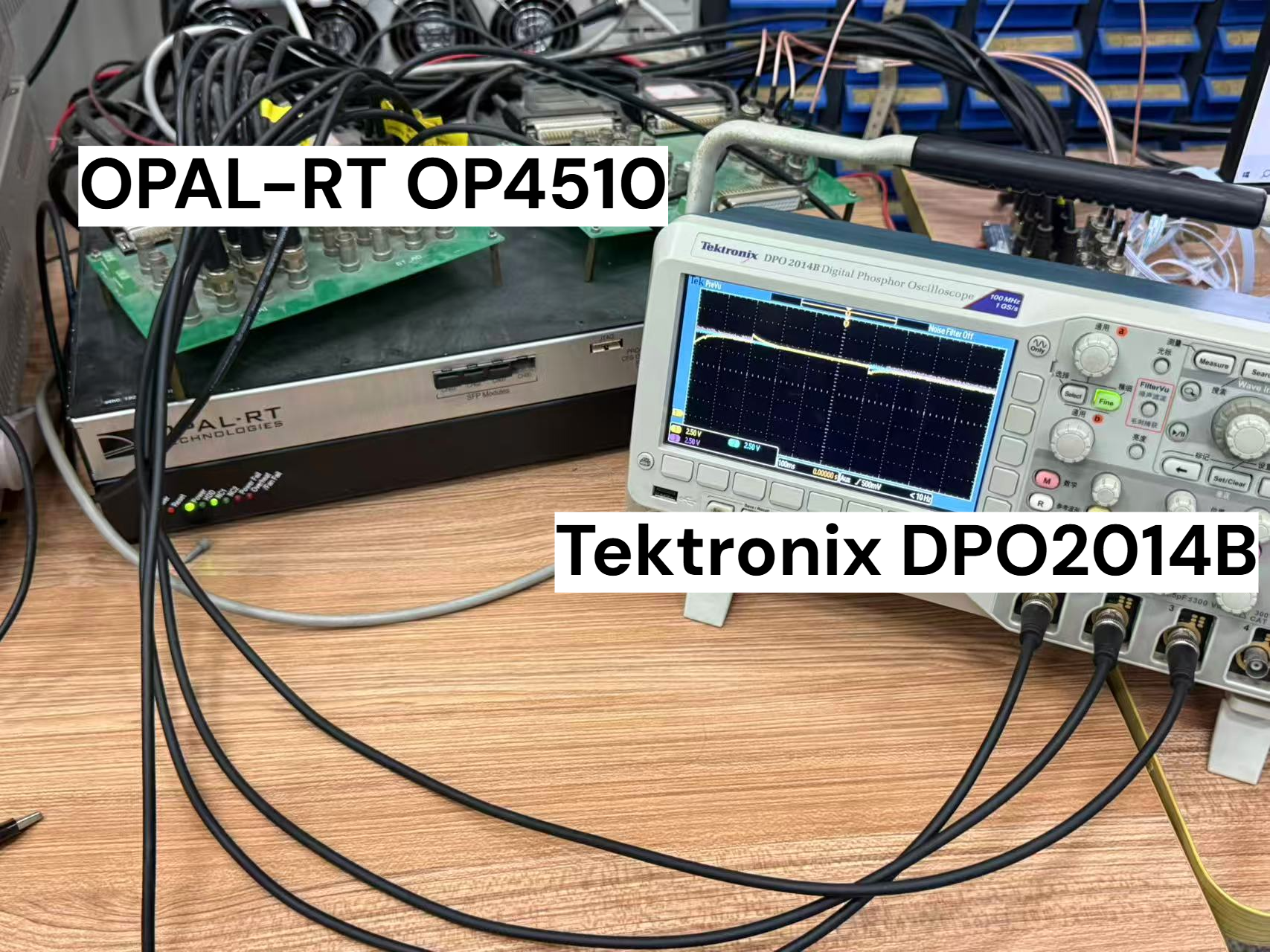}
		\caption{Real-time Simulation Configuration.}
		\label{fig4}
	\end{figure}

	To verify the deployability of the proposed controller, real-time simulation tests were conducted. The simulation platform was the OPAL-RT OP4510 real-time simulator, and the oscilloscope used for signal observation was the Tektronix DPO2014B. The real-time step size was configured to be $1\,\mu\text{s}$. The experimental setup is illustrated in Fig.~\ref{fig4}.
	
	\begin{figure}[!t]
		\centering
		\includegraphics[width=1\linewidth]{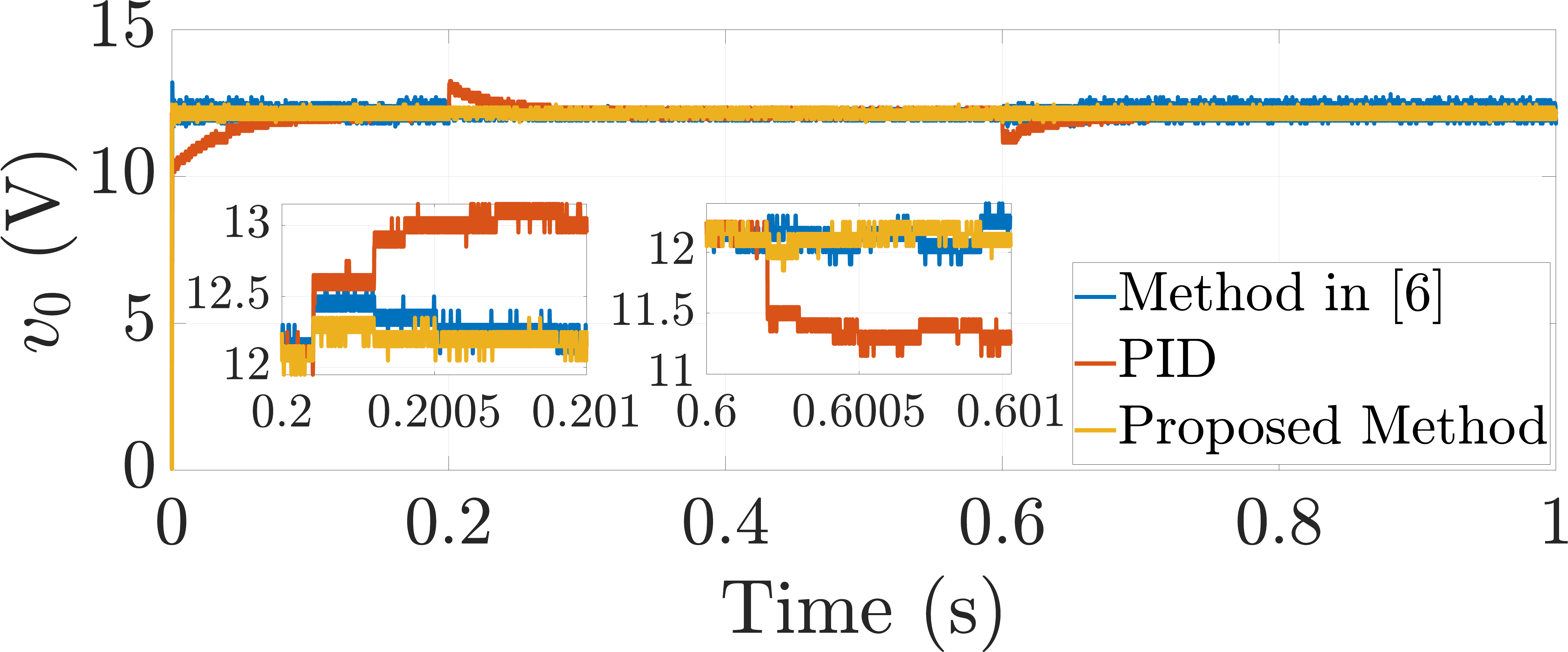}
		\caption{The real-time output voltage response.}
		\label{fig5}
	\end{figure}
	
	\begin{figure}[!t]
		\centering
		\includegraphics[width=1\linewidth]{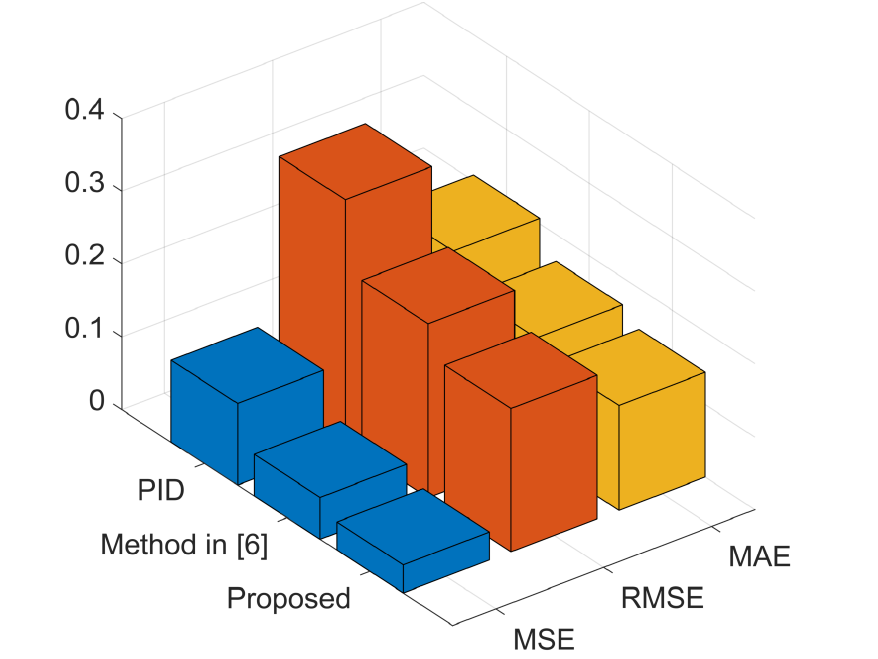}
		\caption{Real-time error metrics.}
		\label{fig6}
	\end{figure}
	
	\begin{table}[htbp]
		\centering
		\caption{Performance comparison of different methods}
		\begin{tabular}{lccc}
			\hline
			\textbf{Method} & \textbf{MSE} & \textbf{RMSE} & \textbf{MAE} \\
			\hline
			Method in \cite{8036261}      & 0.057247  & 0.239264  & 0.162762 \\
			PID                 & 0.112286  & 0.335092  & 0.207601 \\
			Proposed Method     & 0.038693  & 0.196706  & 0.143915 \\
			\hline
		\end{tabular}
		\label{tab2}
	\end{table}

	The real-time simulation results are shown in Fig.~\ref{fig5}–Fig.~\ref{fig6}, and the corresponding numerical values of the error metrics are reported in Table.~\ref{tab2}. By comparing the offline simulation results with those obtained on the OPAL-RT OP4510 platform, it can be observed that the proposed controller is robust to variations in sampling rate, execution delay, and numerical errors, thereby demonstrating strong cross-platform consistency. Moreover, the controller is capable of completing all computations within each control cycle on the OPAL-RT OP4510 platform, which confirms its real-time implementability.
	
	It is also worth noting that slight discrepancies exist between the offline and real-time simulation results. These differences are primarily attributed to: (1) the presence of measurement noise introduced by the oscilloscope, which contaminates the output signals with high-frequency components, and (2) the mismatch in integration time steps between the OPAL-RT OP4510 platform and the Simulink environment running on a conventional PC, which introduces discretization errors.
	
	\section{Conclusion}\label{Sec5}
This paper investigates the fixed-time control problem for boost converter systems. A novel class of functions, termed USSF, is proposed to address the singularity and chattering issues inherent in conventional fixed-time control methods. In addition, a general scaling inequality associated with the USSF is established, which serves as a theoretical foundation for controller design and stability analysis.
Based on the USSF framework, a fixed-time control scheme is developed to ensure that the output voltage converges to a neighborhood of the reference voltage within a fixed time. The proposed controller is rigorously proven to be robust against arbitrarily bounded disturbances. Furthermore, by accepting an increase in system complexity and tuning effort, a disturbance observer based on USSF is designed to further enhance control performance.
The effectiveness and deployability of the proposed control scheme are validated through both non-real-time and real-time simulations. Comparative studies against other control methods demonstrate the superiority of the proposed approach. In the future, we plan to conduct hardware-in-the-loop experiments and deploy the proposed controller on a micro-controller unit to regulate a physical boost converter system.

	\bibliographystyle{ieeetr}
	\bibliography{Reference.bib}
\end{document}